\documentclass[sigconf]{cidr-2027}

\AtBeginDocument{%
  }

\usepackage{todonotes}
\presetkeys{todonotes}{inline}{}
\usepackage{tikz}
\usetikzlibrary{arrows.meta}

\usepackage{graphicx}
\usepackage{subcaption}
\usepackage{multirow}

\usepackage{listings}
\usepackage{xcolor}
\usepackage{verbatim}
\usepackage{url}

\newcommand{\topic}[1]{\par \smallskip \noindent{\bf {#1}}}
\newcommand{\topicul}[1]{\par \smallskip\smallskip \noindent\underline{\bf {#1}}}

\definecolor{codegreen}{rgb}{0,0.6,0}
\definecolor{codegray}{rgb}{0.5,0.5,0.5}
\definecolor{codepurple}{rgb}{0.58,0,0.82}
\definecolor{backcolour}{rgb}{0.95,0.95,0.92}

\lstdefinestyle{mystyle}{
    backgroundcolor=\color{backcolour},   
    commentstyle=\color{codegreen},
    keywordstyle=\color{magenta},
    numberstyle=\tiny\color{codegray},
    stringstyle=\color{codepurple},
    basicstyle=\ttfamily\footnotesize,
    breakatwhitespace=false,         
    breaklines=true,                 
    captionpos=b,                    
    keepspaces=true,                 
    numbers=left,                    
    numbersep=5pt,                  
    showspaces=false,                
    showstringspaces=false,
    showtabs=false,                  
    tabsize=2
}

\usepackage{soul}
\usepackage{todonotes}
\presetkeys{todonotes}{inline}{}

\usepackage[draft,authormarkup=none,commandnameprefix=ifneeded]{changes}
\definechangesauthor[name={Codex},color=blue]{C}
\setdeletedmarkup{\textcolor{red}{\sout{#1}}}
\AtBeginDocument{\ifbool{Changes@optiondraft}{\sloppy}{}}

\begin{document}

\title{When the Application Layer Is Disposable: Putting Integrity Constraints Back in the Database}

\author{Hasara Kalumin}
\affiliation{%
  \institution{University of Maryland at College Park}
  \country{USA}
}
\email{hdesilva@umd.edu}

\author{Amol Deshpande}
\affiliation{%
  \institution{University of Maryland at College Park}
  \country{USA}
}
\email{amol@umd.edu}


\begin{abstract}
Integrity constraints have been a key part of the relational model since the beginning.
However, outside of keys, single-row CHECKs, and referential integrity, most other constraints
are handled by the application layers on top today. SQL ASSERTIONs, standardized in SQL-92, 
remain poorly supported by most major RDBMSs. This has become increasingly unacceptable 
due to two major developments. First, LLM-driven applications/agents increasingly issue
query and update statements directly against the databases, often bypassing the application layer where 
the invariants live today. Such code is envisioned to be generated on demand and discarded afterwards, and as such, the database becomes the only durable component that can encode and enforce 
them. Second, natural constraints in higher-level conceptual models such as the E/R model and property graphs get
fragmented across relations under the physical mappings chosen for performance; the resulting multi-relation 
constraints are not directly supported by most databases and only some are efficiently enforceable by
mechanisms (like triggers) available today. 

These developments lead us to argue for an overhaul of how constraints are defined, mined, managed, and enforced 
in data management systems.
As a first concrete step, we present \emph{constraint indexes}, which
represent constraints spanning multiple relations as transactional access
paths, and implement a proof of concept as a PostgreSQL extension. 
Constraint indexes enforce constraints arising from
distributed inheritance hierarchies, folded relationships, and folded
weak entities, as well as cardinality bounds and functional dependencies,
at low overheads relative to no enforcement. Across these workloads,
they outperform the better of the correctness-preserving 
baselines by up to $2.41\times$. We also show that LLM-driven writers violate
such unenforced constraints at high rates, and behave erratically when 
provided with a list of constraints to enforce.
We close with a discussion of a few research challenges this agenda raises. 
\end{abstract}

\maketitle

\fancyhead[LE,RO]{}

\section{Introduction}
\label{sec:introduction}
\label{sec:intro}

Integrity constraints have been integral to the relational model since its inception,
and were intended to be an intrinsic part of a relation's meaning\footnote{``Integrity constraints specific to a particular relational data base must be definable in the relational data sublanguage and storable in the catalog, not in the application programs.'' (Ted Codd's Rule 10)}.
However, although the SQL standard has supported arbitrary assertions through the
{\tt CREATE ASSERTION} command since 1992, most database systems today support a
limited set of built-in constraints. 
To our knowledge, Oracle 26ai is the only major database system that supports
{\tt CREATE ASSERTION}, but that too comes with significant
limitations\footnote{\url{https://docs.oracle.com/en/database/oracle/oracle-database/26/sqlrf/create-assertion.html}}.

This lack of support primarily stems from the high and unpredictable costs of
enforcing complex constraints, especially those that span multiple tables,
or even multiple rows in a single table. Constraint enforcement also has implications for 
concurrency control, and the semantics of updates under lower isolation 
levels like {\em read committed} become harder to convey. Although it is possible to implement
complex constraints using triggers or shadow tables, over time, enforcement of such 
constraints naturally migrated up the stack to the application layer, where it largely lives
today.

We argue that this is no longer a reasonable division of responsibility, and
that databases must take a far more active role in identifying and enforcing
multi-table constraints. Two developments motivate this argument.
First, data is increasingly modeled and accessed through higher-level abstractions that sit
above the relational model. Commonly used relational mappings transform constraints
that are naturally local in these higher-level models into invariants
spanning multiple relations or involving values embedded within folded
structures, placing them beyond the constraints directly supported by most
databases.
Second, the application layer is no longer the sole place where updates
originate. LLM-driven applications and autonomous agents increasingly
generate database updates directly, often without generating durable
application code in which such constraints can be maintained. Our controlled
study of LLM-generated SQL shows that even supplying the constraints
explicitly does not reliably produce correct enforcement, especially under
concurrent execution. The database must therefore provide both a durable
specification of these constraints and enforcement that is independent of
the behavior of individual writers. Standardized middleware, that {\em all} writes
are required to go through, could also provide a 
durable interface for such writers; that only shifts where our proposed
components live, but doesn't affect the rest of our argument.

In this paper, we present an architecture centered around a
\emph{constraint governor}, envisioned as the central component for
managing the lifecycle of constraints, from their declaration or discovery
through validation, enforcement, revision, and eventual removal.
Constraints may be supplied as part of a conceptual schema or mined from
application code, workloads, or data. The governor keeps track of these
constraints and their status, and decides how each is handled: by which
mechanism, at which enforcement point, with which failure action, and
with what feedback to the writer. These decisions may change over time.
For example, a candidate constraint may initially be checked provisionally
against live traffic, promoted to blocking enforcement as evidence
supports it, or revised or dropped as application requirements change.
A compiler would translate the governor's decisions into the structures and
operations needed for enforcement, including their maintenance,
synchronization, and recovery.

As a first concrete step toward realizing this vision, we present \emph{constraint indexes}, a
database mechanism that represents supported conceptual constraints through mapping-specific, transactionally maintained access paths.
A constraint
index materializes the compact auxiliary state needed to detect
violations efficiently, avoiding repeated multi-table validation queries.
The approach brings together the index structures, maintenance
operations, and synchronization required for constraints that span
multiple relations or involve folded structures.
We implement a proof of concept as a PostgreSQL extension, with
hand-written structures and operations for the evaluated constraints.
The extension stores the auxiliary state in shared memory
and integrates its maintenance
with transaction processing through fine-grained logical locking and
a unified per-transaction undo journal. Updates to the constraint
indexes are atomic with the corresponding relational updates, while
transactions that conflict on constrained values are serialized.
Our evaluation covers constraints arising from distributed inheritance
hierarchies, folded relationships, and folded weak entities, as well as
cardinality bounds and functional dependencies. Across these workloads,
constraint indexes provide efficient validation and scalability under
concurrent execution, outperforming the best baselines 
by up to $2.41\times$
at 256 concurrent clients. These results provide evidence that
specialized transactional access paths can make a substantial class
of nonlocal constraints practical to enforce.

The rest of the paper is organized as follows.
Section~2 elaborates on the two motivating scenarios.
Section~\ref{sec:background} reviews existing enforcement mechanisms
and related work.
Section~\ref{sec:architecture} presents the constraint-governor
architecture and develops constraint indexes.
Sections~\ref{sec:prototype} and~\ref{sec:evaluation} describe the
PostgreSQL implementation and experimental evaluation, respectively.
Section~\ref{sec:conclusion} concludes with research challenges.

\section{Motivation and Problem Setting}
We begin by elaborating on the two motivating scenarios that drive this work.

\subsection{Constraints from Higher-level Conceptual Models}
\label{sec:Higher-level-Conceptual-Models}

As the volume and complexity of data continue to increase, there is a strong case for
elevating the data model abstraction presented to users as well as to AI
agents~\cite{erbium}. Higher-level conceptual models such as the entity--relationship (E/R)
model and the property-graph model~\cite{sun2015sqlgraph,pg-keys} map
more naturally to the real world and are easier to understand, query, and verify.
Real-world constraints also often map naturally to local constraints against such
conceptual models, including uniqueness of attributes across an entity hierarchy,
participation or cardinality constraints, and disjointness across subtypes.
When such schemas are ``shredded'' into relations in different ways depending on
performance goals, the same conceptual constraint fragments differently under each
mapping. Some fragments land on a single relation and can be enforced today by the
underlying database; but many span several tables (or several tuples in a single table) and are not.

\begin{figure}[t]
\hspace{-10pt}
\includegraphics[width=1.05\linewidth]{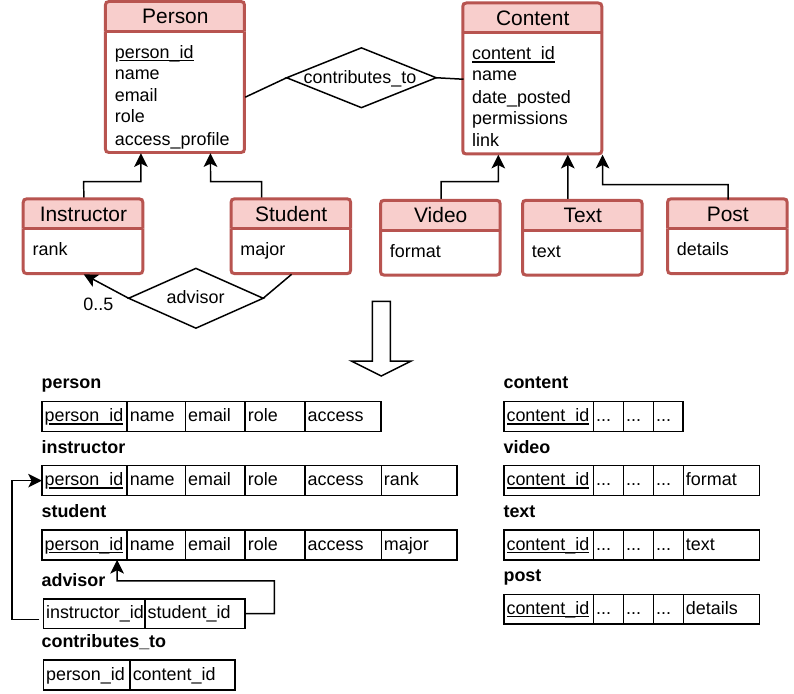}
\vspace{-10pt}
\caption{An example E/R diagram with two hierarchies and one relational mapping for it}
\vspace{-10pt}
\label{fig:er_example}
\end{figure}

\topicul{Entity-Relationship Model.}
Consider the E/R schema in Figure~\ref{fig:er_example} under a {\em concrete-table
mapping}: each instance in the \emph{Person} hierarchy is stored in exactly one of
the \texttt{person}, \texttt{student}, or \texttt{instructor} relations, while each
instance in the \emph{Content} hierarchy is stored in exactly one of the
\texttt{content}, \texttt{video}, \texttt{text}, or \texttt{post} relations. Each
subtype relation repeats the attributes inherited from its parent. This example
illustrates several types of conceptual constraints and how their enforcement
changes under relational mappings:

\topic{Entity key and attribute uniqueness.}
Entity keys ({\em person\_id} and {\em content\_id}), by definition, must be unique across all
entity sets in the respective hierarchy. These natural constraints correspond to multi-table
constraints (spanning 3 tables for {\em person\_id} and 4 tables for {\em content\_id}) in the 
relational mapping above. Similarly, if {\em email} is declared unique in the {\em Person} entity set, that uniqueness must be enforced 
across the \texttt{person}, \texttt{student}, and \texttt{instructor} relations.
Such \texttt{unique} constraints are not directly specifiable in most databases today.

\topic{Referential integrity.} 
Referential integrity constraints are typically enforced through the {\tt FOREIGN KEY} construct. However,
for a relationship defined at higher levels of a hierarchy, there may be no single target table for that construct.
For 
\texttt{contributes\_to}, the person endpoint may refer to a tuple in
\texttt{person}, \texttt{student}, or \texttt{instructor}, while the content
    endpoint may refer to a tuple in \texttt{content}, \texttt{video},
    \texttt{text}, or \texttt{post}. 

\topic{Participation and cardinality constraints.}
    Participation constraints specify whether an entity must take part in a
    relationship, while cardinality constraints bound the number of
    relationship instances in which it may participate. For example, the
    \emph{advisor} relationship restricts each instructor to at most five
    advisees. When the relationship is stored as a separate relation, enforcing
    this bound requires counting multiple relationship tuples. Such aggregate
    bounds cannot be expressed directly using ordinary keys, foreign keys, or
    row-level \texttt{CHECK} constraints. Similarly, a foreign key validates
    existing relationship endpoints but does not ensure that every entity
    participates in a required relationship.

\topic{Functional dependencies beyond keys.}
    Applications may also require a set of attributes to determine another set
    without uniquely identifying an entity. For example,
    \(\mathit{role} \rightarrow \mathit{access\_profile}\) requires all persons
    with the same role to have the same access profile. A \texttt{UNIQUE}
    constraint on \emph{role} would be too restrictive because multiple persons
    may share a role. Enforcing the dependency requires comparing values across
    entities, potentially spanning multiple relations when the hierarchy is
    distributed.

\topic{Weak-entity identity and ownership (not shown in the example).}
A weak entity is identified by its owner's key together with a partial key
and cannot exist without its owner. If the weak entity is stored as a separate table, these
conditions can often be expressed using a composite primary key and a
foreign key to the owner. But if weak entities are folded into owner relations
as arrays of composite values, their identifying and other constrained
attributes are embedded within array elements. The standard \texttt{PRIMARY KEY},
\texttt{UNIQUE}, and \texttt{FOREIGN KEY} declarations do not directly enforce
constraints on those elements. Enforcement must instead examine the folded
values, including values in other owner tuples when an attribute is required
to be globally unique.

Some of these issues can be seen in existing object-relational mappers and how they
handle constraints defined at the conceptual
level~\cite{jakarta_unique_constraint,
hibernate_adding_constraints,django_field_unique,
sqlalchemy_concrete_inheritance,
rails_uniqueness_validation}. For instance, SQLAlchemy describes concrete inheritance,
as in the example above, as substantially more complicated, particularly
for relationships and polymorphic loading. Single-table inheritance, on the other hand, cannot easily handle
{\tt NOT NULL} constraints on subtype attributes, and those constraints may be
dropped, although multi-attribute {\tt CHECK}s may work in some cases.

\topicul{Property Graph Model.}
The same pattern can be seen in some of the recent property-graph schema proposals.
PG-Keys~\cite{pg-keys} provides a general model for defining keys over
property graphs. A PG-Key uses one query to define its \emph{scope}, i.e., objects to 
which the PG-Key applies, and another to define its \emph{descriptor}, i.e.,
how to obtain key values associated with each object. Both queries may depend on
object properties and graph topology, with their expressiveness determined
by the chosen query language.
A PG-Key may make a few different assertions: \emph{exclusive}
means that no two objects in the scope may share a key value;
\emph{mandatory} means that every object in the scope must have at least one key
value; and \emph{singleton} means that every object in the scope may have at most one
key value. An \emph{identifier} combines all three requirements.

Some such constraints map directly to conventional relational
constraints; for example, uniqueness of a node property ({\em exclusive}) may become a
\texttt{UNIQUE} constraint on the corresponding relation, and {\em mandatory} assertions
may map to \texttt{NOT NULL} constraints.
More complex scopes and descriptors, however, may require enforcement
across relations. For example, a PG-Key may express the constraint that
every forum with members must have exactly one moderator (in a {\em social network} domain), 
and that moderator must uniquely identify the forum among forums with members~\cite{pg-keys}.
Under a natural mapping that stores \texttt{hasMember} and
\texttt{hasModerator} edges in separate relations, membership determines
which forums participate in a uniqueness constraint on moderator
assignments. Adding a forum's first member can therefore introduce a
uniqueness violation without changing any moderator assignment.
A \texttt{UNIQUE} constraint on the moderator identifier in the moderator relation would impose a
stronger condition, also restricting forums without members.
Enforcing such constraints requires accounting for changes to multiple
relations, motivating the shared auxiliary state and transactional
enforcement techniques developed in this paper.

\subsection{LLM-Driven Writers and Constraints}

At the same time, the application layer is no longer the sole or even primary place
where updates originate. LLM-driven applications and autonomous agents increasingly
issue statements directly against databases, often purely based on the schema,
and the code they run is frequently generated for a single task and discarded. If
invariants are encoded in such code, they are unreviewed and transient, and do not
typically accumulate over time.
As such, the database is the only component of the stack that persists across all such writers,
which also makes it the only place where a durable specification of valid data can be
kept, as originally intended, and the last point at which a violating write can be
deterministically stopped.

To obtain concrete evidence for the claim that LLMs cannot be trusted to maintain
integrity constraints that are not encoded in the schema, we ran a small probe against
the schema above, with the \texttt{content} table dropped (full specialization). We selected six such constraints: \\
(a) two governing the
shared ID/name space across the 
\{\texttt{contentpost}, \texttt{contenttext}, \texttt{contentvideo}\} specialization, \\
(b) two
governing the shared ID/email space across
\{\texttt{person}, \texttt{student}, \texttt{instructor}\}, \\
(c) advisor fan-out cap of at
most five advisees per instructor, and  \\
(d) the functional dependency \texttt{role} $\rightarrow$ \texttt{access\_profile}. \\
For each constraint, we constructed one concrete, fully specified update or insert
scenario---every attribute was given a literal value except the one governed by the
constraint---and asked two models, Claude Sonnet~5 and GPT-5.6-Terra, to write the SQL
for it five times each, first with no mention of the constraints and then with all six
stated explicitly in the prompt.
We inspected and classified the generated SQL; the concurrency observations below identify
potential violations from the checking and locking strategies in the code.

Without the constraints stated, the result was unambiguous: in all 60 trials, the
returned SQL applied no check whatsoever---not even the single-table primary-key
uniqueness; although the schema itself never declares that explicitly, those are natural and common constraints.
Every response was a bare
\texttt{INSERT} or \texttt{UPDATE} that, if executed, would silently violate the
intended invariant.

Given the constraint list explicitly in the prompt, behavior became far more varied, and the range
of failure modes is itself instructive:

\topicul{(1) {Silent restriction (40/60 trials):}}
The dominant response added a
\texttt{WHERE NOT EXISTS(\ldots)} or
\texttt{(SELECT COUNT(*)\ldots) < 5} condition around the write.
If the condition is not true, the statement simply affects zero rows,
without raising an error or providing constraint-specific feedback.
The caller must inspect the affected-row count to detect that no change was made.
In several cases, the model's output flagged the danger but did
not fix it in the code.

\topicul{(2) {Check-then-raise (15/60 trials):}}
Other responses wrapped the write in a \texttt{DO} block that re-runs the same checks
and \texttt{RAISE}es \texttt{EXCEPTION}s on violation. Under PostgreSQL's default
\texttt{READ COMMITTED} (and, short of true \texttt{SERIALIZABLE}, under snapshot
isolation more generally), a bare check-then-act is a textbook write-skew race.
Only one of the 60 trials used PostgreSQL’s general solution, \texttt{SERIALIZABLE} with retries. 
Within the check-then-raise category, some responses used explicit locks,
with differing implications for correctness and concurrency.
\texttt{LOCK TABLE \ldots IN SHARE ROW EXCLUSIVE MODE} blocks \emph{every} writer to
that table, because an ordinary \texttt{INSERT} or \texttt{UPDATE} requires
\texttt{ROW EXCLUSIVE}, which conflicts with it. It therefore closes the race, but at
the cost of serializing all writers to every locked table, including, in the most
extreme trial, seven of the schema's eight tables to insert one content row.
\texttt{SELECT \ldots FOR UPDATE} on a proxy row, such as locking the
\emph{instructor} row to guard a count taken over the \emph{student} table, and
\texttt{pg\_advisory\_xact\_lock} calls are weaker. They serialize only against
other transactions that adopt the identical locking convention and do nothing
against a plain, unguarded \texttt{UPDATE}, which is exactly what a sibling trial of
the same advisor-cap scenario produced.

\topicul{(3) {No guard at all (4/60 trials):}}
Even with every constraint spelled 
out, a handful of responses were
indistinguishable from the no constraints baseline.

The remaining trial produced an empty response. This inspection shows
that supplying constraints as text does not reliably produce correct
enforcement. Stating the constraints prompts more checks, but also much
more complex and varied SQL, leaving each generated statement responsible
for choosing and correctly implementing an enforcement and
concurrency-control strategy.
For instance, for the {\em person\_id} constraint, the generated SQL was on average 153 words long when provided with the
constraints to enforce, compared to just 16 words with no constraints.
In several cases, the generated SQL had placeholders without any warnings, which would have corrupted the
database silently as well.
This further motivates that the database must provide durable, machine-readable constraints and enforce them
independently of the behavior of any individual writer.

\section{Background}
\label{sec:background}

Relational integrity constraints range from restrictions on individual
attribute values to conditions over multiple tuples and
relations~\cite{turker2001semantic}. 
Most databases today support a limited set of constraints that can be specified when
creating or altering tables, or through explicit constraint DDL statements. These
include primary keys, candidate keys ({\tt UNIQUE}), domain constraints
({\tt CHECK}, {\tt NOT NULL}), and referential integrity ({\tt FOREIGN KEY}).
Although the SQL standard allows a {\tt CHECK} constraint to include subqueries (e.g., to compare against other rows), most databases, including PostgreSQL, restrict {\tt CHECK} to operate on one row.
Other constraint classes commonly discussed in the literature, including general
functional dependencies, participation constraints for many-to-many relationships, inclusion dependencies that are not foreign keys, and denial
constraints~\cite{chu2013denial}, are rarely supported directly. 
More
general assertions, supported in the standard through {\tt CREATE ASSERTION}, have
only limited support.

The only multi-row conditions relational systems typically enforce efficiently are
uniqueness and referential integrity, which are not checked by re-evaluating a query but utilize
a B+-tree on the relevant attribute instead. 
PostgreSQL's \texttt{EXCLUDE}
constraints generalize this idea by enforcing that no two rows satisfy a given
operator relationship, most usefully non-overlap of ranges, backed by a GiST index
that supplies both the search and the locking. \texttt{EXCLUDE} is, in our terms, a
constraint compiled into an access path. It is confined to a single relation, and
nothing analogous exists across relations.

Given this lack of direct support, two techniques are commonly used to handle the types of multi-relation constraints discussed above.

\subsection{Triggers}
\label{sec:triggers}

A direct approach attaches triggers to the participating relations.
For each relevant modification, a trigger queries the other relations for
a violation and raises an error if one is found.
For example, to enforce uniqueness of \texttt{person\_id} and \texttt{email}
across \texttt{person}, \texttt{student}, and \texttt{instructor} tables, an
insertion or update trigger on each table can check whether either value
already appears in the other two tables. Native uniqueness constraints
handle duplicates within each table.

These checks alone do not guarantee correctness under concurrency.
Under PostgreSQL's default \texttt{READ COMMITTED} isolation level,
two transactions may concurrently insert the same \texttt{person\_id}
into \texttt{student} and \texttt{instructor}. Each trigger's query can
fail to observe the other transaction's uncommitted tuple, allowing both
transactions to commit, violating hierarchy-wide uniqueness.
This is a write-skew anomaly: the transactions modify different relations
while relying on checks over a shared invariant.
Correct enforcement requires consistent use of \texttt{SERIALIZABLE}
isolation for the participating transactions, with retries after serialization
failures, or an explicit locking protocol that serializes operations on
conflicting conceptual keys.

Query-based triggers also incur repeated query execution for relevant writes,
with costs that depend on the number of participating relations and the
available indexes. 
Moreover, triggers are procedural rather than declarative (making it hard to use them as the constraint definition)
and often require custom code for each constraint. Developers must ensure
that these checks cover every relevant modification and maintain the
checking and synchronization logic as the schema, mapping, or supported
operations evolve.

\subsection{Shadow Tables}

\begin{figure}[t]
\centering
\includegraphics[width=\linewidth]{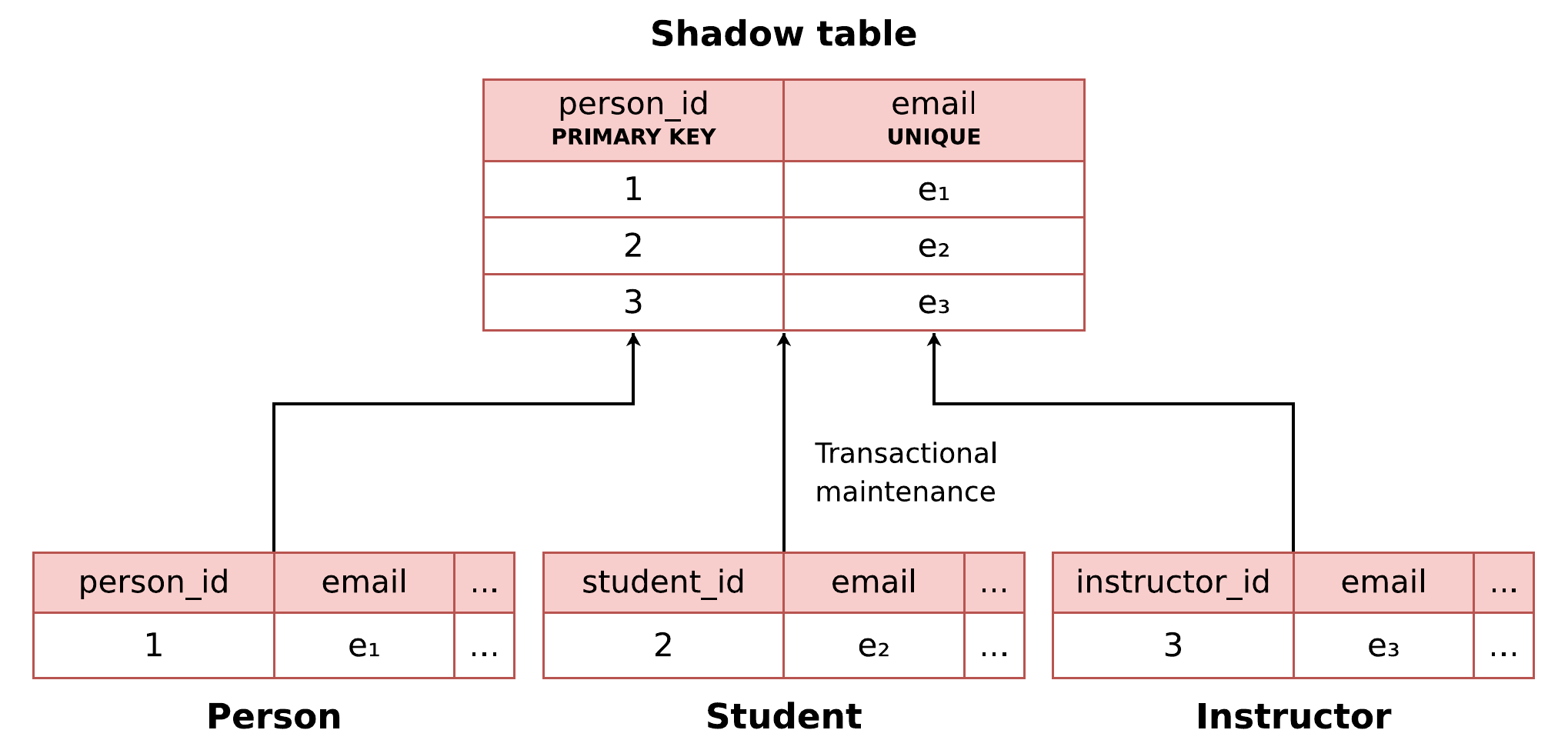}
\caption{A shadow table enforces hierarchy-wide uniqueness through separate
unique indexes on \texttt{person\_id} and \texttt{email}. Each source modification
maintains the corresponding shadow-table row within the same transaction.}
\label{fig:shadow_table_example}
\end{figure}

Another option is to build {\em shadow tables}: {\em materialized views} defined over
multiple source relations so that a multi-relation constraint becomes a local
constraint on the shadow table. 
For example, Figure~\ref{fig:shadow_table_example}
shows a shadow table containing one (\texttt{person\_id}, \texttt{email}) row for each entity stored in
\texttt{Person}, \texttt{Student}, or \texttt{Instructor}. 
The two uniqueness requirements across the hierarchy can now be enforced through a
primary key constraint on \texttt{person\_id} and a separate unique constraint on \texttt{email}.
This approach is correct because the unique indexes perform the locking that the
trigger's snapshot read cannot, and it is reasonably fast because enforcement again
rides on an access path and we avoid the repeated queries over sibling relations. 
Correctness requires every relevant insertion, update, and deletion to
maintain the shadow table within the same transaction.
Several constraints may share a shadow table, as in this example, but
the auxiliary schema, indexes, and maintenance operations must still be
designed to preserve all relevant constraints under concurrency and abort.
Where automatic incremental maintenance is unavailable, triggers or
explicit update procedures are needed to maintain these tables.

Furthermore, adding a constraint requires a schema change and backfill; removing one requires
finding every maintenance site. Shadow tables also conventionally offer only one enforcement
strength: blocking the transaction. There is no easy way to introduce a candidate
constraint provisionally, observe how often live traffic would violate it, and promote
it only if it appears valid. The technique also does not extend directly beyond
uniqueness: cardinality bounds, interval disjointness, and aggregate conditions each
require a different structure designed and maintained from scratch.
PostgreSQL's native materialized views require explicit refreshes and do
not support automatic incremental maintenance. Consequently, they do not
directly provide the transactional maintenance needed for shadow-table
constraint enforcement.

\subsection{Prior Work}
We briefly discuss other related work on integrity constraints.

\noindent\textbf{Constraints in conceptual models.}
Work on conceptual models includes incremental validation of UML/OCL
constraints~\cite{cabot2009incremental,oriol2015ocl} and consistency
checking when model constraints are defined or revised
dynamically~\cite{groher2010dynamic}. PG-Schema~\cite{pg-schemas} combines
type definitions, including multiple inheritance, with constraints based
on PG-Keys, discussed earlier. These approaches provide languages and
validation techniques for conceptual invariants; enforcing those
invariants under different relational mappings is one of our motivating
concerns.

\noindent\textbf{Incremental constraint validation.}
A substantial body of work reduces validation cost by deriving
update-specific checks or compiling declarative constraints into
production and event
rules~\cite{qian1986knowledge,ceri1990deriving,olive1991integrity,
urpi1992change}. TINTIN~\cite{oriol2016tintin} generates auxiliary tables,
triggers, and incremental SQL queries to validate pending changes before
applying them. More recent work extends the underlying event rules to
handle aggregates and improve checks involving existential
conditions~\cite{oriol2025incremental}.
Integrity checking is also closely related to incremental view
maintenance: violations can be represented as query results required to
remain empty. Prior work derives view-maintenance rules and selects
auxiliary views to reduce maintenance and integrity-checking
costs~\cite{ceri1991views,ross1996materialized}.
Cavero Barca et al.~\cite{cavero2019crossrow} implement cross-row assertions
through materialized violation counts constrained to remain zero.
DBToaster and DBSP~\cite{koch2014dbtoaster,budiu2023dbsp} provide more
general techniques for compiling incremental computations.

\noindent\textbf{Enforcement and concurrency.}
T\"urker and Gertz~\cite{turker2001semantic} survey declarative
constraints, assertions, and triggers. Native assertion support, such
as Oracle's \texttt{CREATE ASSERTION}~\cite{oracle2026assertions},
provides another enforcement mechanism.
Efficient validation alone, however, does not ensure correctness under
concurrent updates. Predicate locking~\cite{eswaran1976notions} and
assertion-specific synchronization, as in
UQJG~\cite{koppelaars2021uqjg}, address coordination among transactions
that may jointly violate a constraint. The recent work on incremental
event rules~\cite{oriol2025incremental} identifies concurrent assertion
checking as a direction for further work.

\noindent\textbf{Application constraints and policy management.}
Studies of database-backed applications document difficulties in
maintaining constraints across application components and versions,
and violations caused by concurrent application-level
checks~\cite{yang2020managing,bailis2015feral}.
CFinder~\cite{huang2023protecting} infers constraints from application
code and identifies those missing from the database schema, while
Coco~\cite{liu2023leveraging} extracts constraints for physical design
and query optimization. Such techniques could supply candidate
constraints to a constraint governor.
Ataullah and Tompa~\cite{ataullah2011business} address centralized
business-policy management by translating workflow and lifecycle
restrictions into temporal integrity constraints and reasoning about
policy conflicts.

Our focus is on dynamic constraint governance: choosing how and when
a constraint is enforced, how violations are handled, and how these
choices evolve. Constraint indexes provide one enforcement mechanism
within this framework, combining compact shared auxiliary state with
transactional synchronization for constraints spanning multiple
relations or folded values.

\section{Towards Constraint Governance}
\label{sec:architecture}

\begin{figure}[t]
    \centering
    \includegraphics[scale=0.95]
    {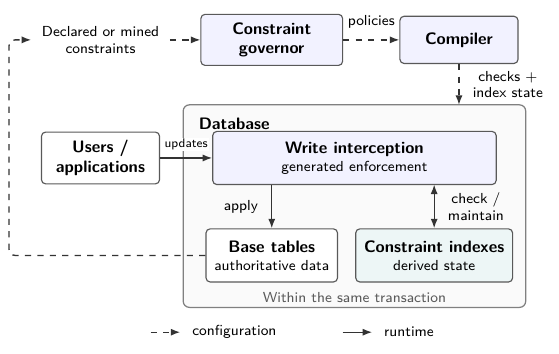}
    \vspace{-15pt}
    \caption{Constraint Governor Envisioned Architecture}
    \vspace{-15pt}
    \label{fig:architecture}
\end{figure}

Motivated by the observations made in the previous section, we begin by sketching an
architecture for making constraint enforcement a centralized, governed activity (Figure~\ref{fig:architecture}). At its center is a \emph{constraint governor}, in charge of deciding
how each constraint is handled: by which mechanism, at which enforcement point, with
which failure action, and with what feedback to the writer. A compiler module would be in charge
of translating these decisions into auxiliary state, corresponding data structures, and/or 
triggers.  We then introduce \emph{constraint indexes}, 
transactional access paths that represent key-oriented multi-relation constraints in shared enforcement structures, 
as one example of such auxiliary state.

\subsection{Constraint Governor}
\label{sec:governor}
We envision constraints entering the system either as schema declarations
or as candidates mined from application code, workloads, or data. The governor
is responsible for keeping track of the active constraints and choosing, for each constraint,
an enforcement mechanism, an
\emph{enforcement point} (per statement, at commit, or asynchronously), a
\emph{failure action} (abort, quarantine, or alert only), and the feedback returned
to the writer. The constraint governor may primarily optimize for performance, but may also
need to consider the impact on overall data consistency; how to navigate the trade-offs between 
these factors is a key open research question here.
A compiler would translate these choices into derived state and generate
the operations, synchronization, undo, and reconstruction logic needed to maintain it.
Writers (applications, LLMs) continue to issue ordinary updates, while the underlying data system 
remains authoritative.

We propose separating the mechanism, enforcement point, and failure action so that
constraints can evolve through a lifecycle. The system may first evaluate a candidate
constraint \emph{provisionally}, raising alerts without rejecting writes, and later
promote it to blocking enforcement, demote it, or drop it. It may also enforce the same
constraint strictly on one branch, provisionally on another, or asynchronously over
historical or analytical data.

We envision two deployment modes. In a transactional database, generated operations
run on the write path, synchronize conflicting transactions, maintain derived state,
and prevent violating transactions from committing. In a disaggregated system such as
a lakehouse, no common transactional point may exist at which the system can intercept
every write. The system can instead copy updates, commit records, or change events to
a constraint monitor, which maintains the same derived state and detects violations
as they arrive. Because the monitor operates outside the committing path, it may alert,
quarantine data, prevent downstream publication, or initiate repair, but cannot
necessarily abort the original write. It can rebuild its state from authoritative
tables or snapshots and then replay subsequent updates.

Because we treat enforcement state as derived, the system can add or drop a constraint
by generating or discarding its structures and rebuilding them from the underlying
data. This avoids adding persistent auxiliary relations to the schema,
although initializing a new constraint still requires deriving its state
from existing data.
It can also move a compiled constraint between monitoring
and blocking modes without requiring a new hand-written implementation. 

\subsection{Constraint Indexes}
A \emph{constraint index} represents compact auxiliary state sufficient to
enforce a supported constraint. The compiler would derive this state, store that state in a shared auxiliary organization, and generate the operations and logical locks
required to maintain it transactionally. The auxiliary state does not replicate
complete relational tuples; each auxiliary object contains only the identities,
unique values, and ownership information required for enforcement. It could also 
contain custom data structures like Bloom filters.

Consider the disjoint \texttt{Person}--\texttt{Student}--\texttt{Instructor}
hierarchy from \S\ref{sec:Higher-level-Conceptual-Models}. Although
instances are stored in three relations, they share one conceptual identity
space, and \texttt{email} is unique across the complete hierarchy. The compiler
would therefore create a compact auxiliary object:
\[
o_{\mathsf{Person}}
=
\langle
\mathit{person\_id},
\mathit{email}
\rangle,
\]
and two indexes that reference it:
\[
I_{\mathsf{Person}}[\mathit{person\_id}]
\rightarrow
o_{\mathsf{Person}}^{*},
\qquad
U_{\mathsf{email}}[\mathit{email}]
\rightarrow
o_{\mathsf{Person}}^{*}.
\]
Every insertion into \texttt{Person}, \texttt{Student}, or
\texttt{Instructor} reserves its identifier in
$I_{\mathsf{Person}}$, so transactions modifying different physical relations
still conflict when they use the same conceptual identity. If an email is
present, the insertion also reserves it in $U_{\mathsf{email}}$. Because both
indexes reference the same auxiliary object, an email update changes only the
email index entry and the corresponding field in the object, while a deletion
can locate and remove all associated entries directly.

The same principle applies to folded objects. For a weak entity
\texttt{IdCard(card\_id, card\_no)} owned by \texttt{Person}, the enforcement-relevant state is
\[
o_{\mathsf{IdCard}}
=
\langle
\mathit{person\_id},
\mathit{card\_id},
\mathit{card\_no}
\rangle.
\]
The compiler would generate an identity index for the composite weak-entity key,
a secondary index for the unique card number, and an owner index:
\[
I_{\mathsf{IdCard}}
[(\mathit{person\_id},\mathit{card\_id})]
\rightarrow
o_{\mathsf{IdCard}}^{*},
\]
\[
U_{\mathsf{card\_no}}[\mathit{card\_no}]
\rightarrow
o_{\mathsf{IdCard}}^{*},
\]
\[
O_{\mathsf{IdCard},\mathsf{Person}}
[\mathit{person\_id}]
\rightarrow
\{o_{\mathsf{IdCard}}^{*}\}.
\]
The first two indexes enforce weak-entity identity and
\texttt{card\_no} uniqueness even when \texttt{IdCard} values are embedded in
owner tuples distributed across several relations. The owner index groups all
dependent objects by owner, allowing owner deletion and cascading maintenance
without scanning or extracting every folded value. Thus, a constraint index
consolidates the enforcement-relevant state into shared lookup paths,
while the relational representation remains the complete and durable source of
application data.

\subsection{Compilation}
For each supported constraint domain $\Gamma$, we envision a compiler
that produces
\[
\mathcal I_{\Gamma}
=
\bigl(
S_{\Gamma},
\mathsf{Ops}_{\Gamma},
L_{\Gamma},
U_{\Gamma},
B_{\Gamma}
\bigr).
\]
This specifies the intended compilation output. The current prototype
implements these components by hand for the evaluated constraints, using
a set of constraint-specific templates; automatic compilation
from a conceptual schema and relational mapping remains future work.
The five components are as follows:
$S_{\Gamma}$ is the auxiliary state used to validate the
constraints in $\Gamma$, including the generated identity, unique-value,
and owner-index entries.
$\mathsf{Ops}_{\Gamma}$ is the set of generated operations for
accessing and maintaining this state, such as reservation, update, and removal.
$L_{\Gamma}$ specifies the logical keys and lock modes that each
operation must acquire so that conflicting transactions synchronize even
when they modify different physical relations.
$U_{\Gamma}$ specifies the inverse actions used to undo speculative
auxiliary-state changes when a transaction aborts.
$B_{\Gamma}$ is the reconstruction procedure that rebuilds the
auxiliary state $S_{\Gamma}$ from the durable relational database after startup or
recovery.

For example, the constraint domain that enforces hierarchy-wide identity
and email uniqueness for \texttt{Person} maintains the auxiliary state
\[
S_{\mathsf{Person}}
=
\bigl(
I_{\mathsf{Person}},
U_{\mathsf{email}},
\{o_{\mathsf{Person}}\}
\bigr),
\]
where
$I_{\mathsf{Person}}$ indexes auxiliary objects by
$\mathit{person\_id}$,
$U_{\mathsf{email}}$ indexes the same objects by email, and
$\{o_{\mathsf{Person}}\}$ denotes the set of auxiliary objects
\[
o_{\mathsf{Person}}
=
\langle
\mathit{person\_id},
\mathit{email}
\rangle .
\]
The generated operations
$\mathsf{Ops}_{\mathsf{Person}}$ include looking up an identifier, reserving a
new identifier and email, changing an email, and removing an object. For
example, inserting a person with identifier $k$ and email $u$ reserves entries
\[
I_{\mathsf{Person}}[k]\rightarrow o^*,
\qquad
U_{\mathsf{email}}[u]\rightarrow o^*.
\]
The lock specification $L_{\mathsf{Person}}$ maps this insertion to exclusive
logical locks on the corresponding identity and email keys:
\[
L_{\mathsf{Person}}(\mathsf{insert}(k,u))
=
\{X(\mathsf{id},k),X(\mathsf{email},u)\}.
\]
Consequently, insertions into different physical hierarchy relations conflict
whenever they use the same conceptual identifier or email.
The undo specification $U_{\mathsf{Person}}$ records the inverse of each
speculative change. If the insertion aborts, it removes the entries for $k$
and $u$ and frees the newly created auxiliary object. For an email update from
$u_{\mathit{old}}$ to $u_{\mathit{new}}$, it restores the old email entry and
the old value in the auxiliary object.
Finally, $B_{\mathsf{Person}}$ scans the durable
\texttt{Person}, \texttt{Student}, and \texttt{Instructor} relations and
reconstructs the auxiliary objects and both indexes after startup or recovery.

\subsection{Transactional Maintenance}
Each update using the constraint-index operations follows
\[
\mathsf{Acquire}\rightarrow\mathsf{Validate}\rightarrow
\mathsf{Reserve}\rightarrow\mathsf{Update}\rightarrow\mathsf{CommitOrUndo}.
\]
It acquires transaction-duration locks on the affected conceptual keys, validates and
speculatively updates the auxiliary state, and then performs the relational update.
On commit the changes stand; on abort, $U_{\Gamma}$ restores the prior state before
the locks are released.

Writing $\Phi_{\Gamma}$ for the function that derives auxiliary state
from committed relational data $D$, correctness requires
$S_{\Gamma}=\Phi_{\Gamma}(D)$ at quiescent points, when no transaction
has tentative changes in the shared structures. During concurrent
execution, these structures also contain tentative reservations protected
by logical locks. The maintenance protocol must ensure that
all constraint-relevant writes use the constraint-index operations, every pair of logically 
conflicting operations acquires a common key under incompatible lock modes, and every
speculative change has an inverse. 
A monitoring-only deployment would instead process an observed update
stream and report violations. Its auxiliary state must also represent
violating data, such as multiple objects sharing a supposedly unique value;
adapting the enforcement structures to this setting remains future work.

\begin{figure*}[t]
    \centering
    \includegraphics[scale=0.35]
    {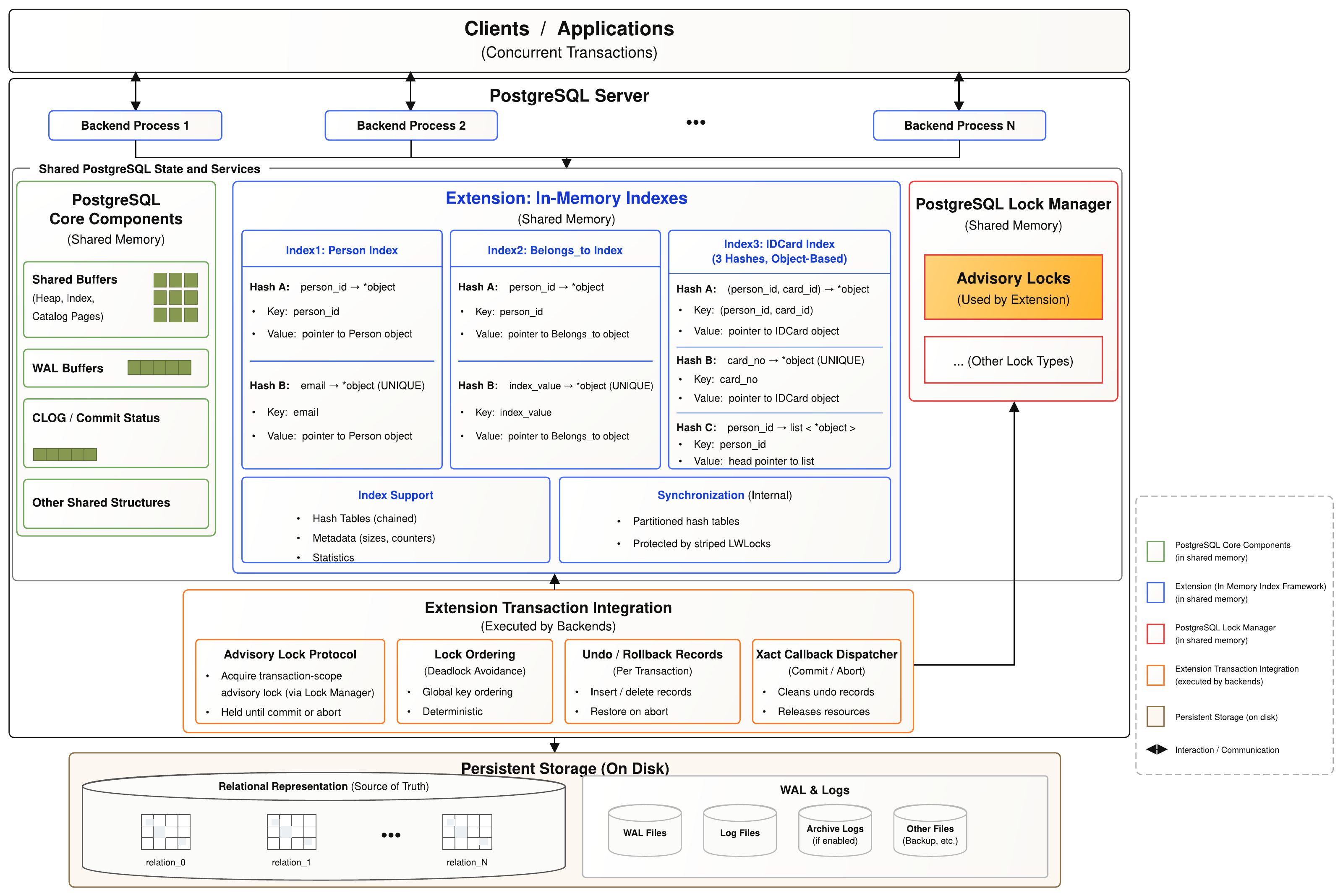}
    \caption{PostgreSQL constraint-index architecture. Shared-memory indexes
    store derived constraint state, advisory locks synchronize conceptual
    keys, and transaction callbacks provide commit/abort processing. The
    relational representation remains the durable source of truth.}
    \label{fig:prototype-architecture}
\end{figure*}

\section{PostgreSQL Prototype}
\label{sec:prototype}

We implemented constraint indexes as a PostgreSQL extension loaded at server
startup through \texttt{shared\_preload\_libraries}, allowing the extension
to reserve and initialize server-wide shared memory. Figure~\ref{fig:prototype-architecture}
shows how the extension integrates with PostgreSQL backends, the lock manager,
transaction processing, and persistent storage.
The extension maintains its derived constraint state in partitioned
shared-memory hash tables protected internally by striped lightweight locks.
When executing an update, a backend acquires transaction-scoped advisory locks
for the affected constrained values in a deterministic order, modifies the
relevant constraint indexes, and appends inverse operations to a unified
per-transaction undo journal. PostgreSQL transaction callbacks discard the
journal on commit and replay it in reverse order on abort. The constraint
indexes are derived and non-durable; the relational representation
remains the source of truth from which they can be rebuilt after a server
restart. We elaborate in the rest of the section.

We note that the current prototype is a proof of concept for the enforcement
mechanism. Specifically, it assumes that all relevant writes invoke the hand-written
constraint-index operations in the same transaction as the relational
updates; it does not intercept or prevent direct SQL writes that bypass
these operations. 

\begin{figure}[t]
    \centering
    \includegraphics[scale=0.35,
    trim=0cm 2.5cm 0cm 0cm,
    clip]{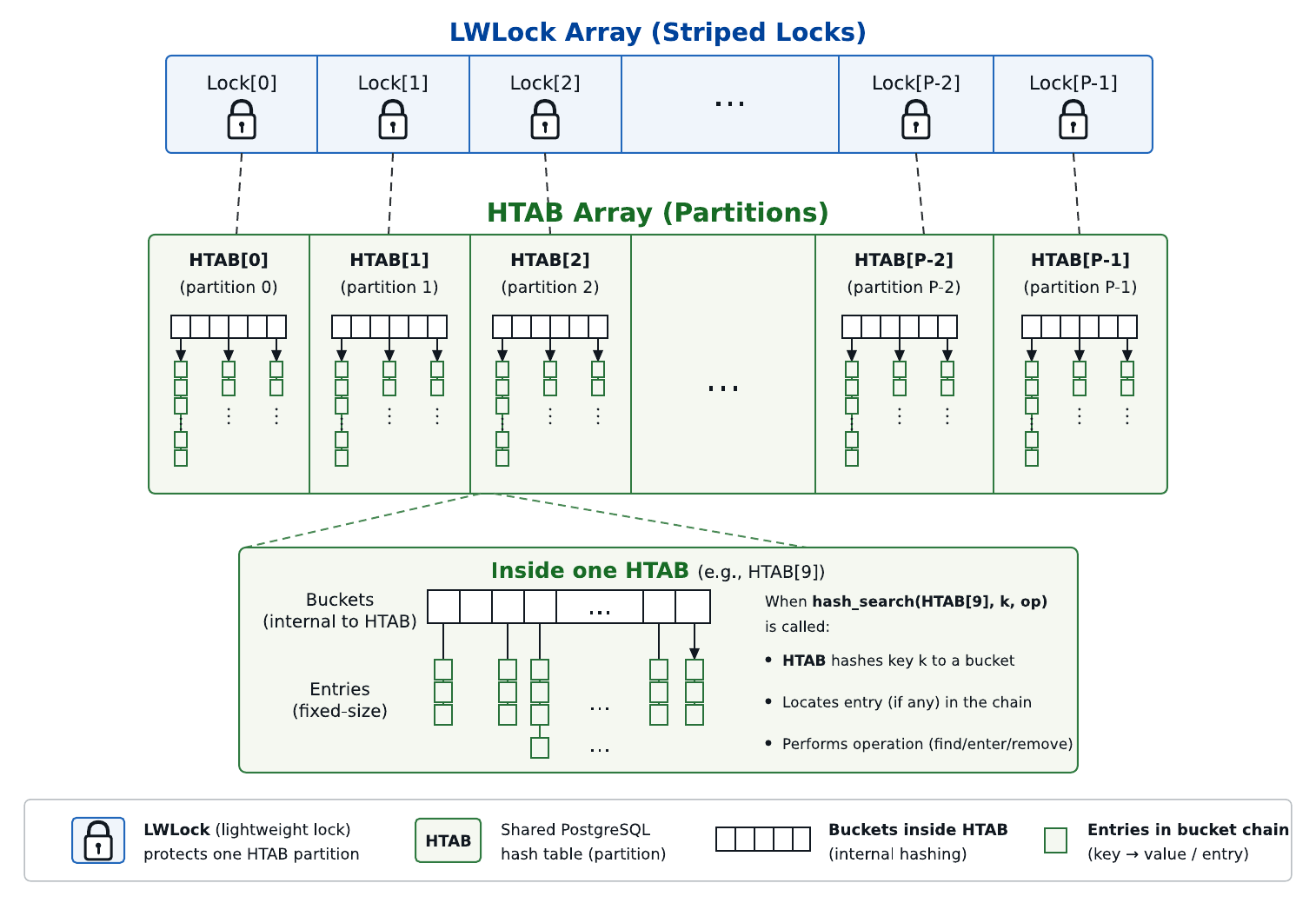}
    \caption{Partitioned implementation of one shared hash structure. An
    operation maps $k$ to partition $p$, acquires
    \texttt{LWLock}$[p]$, and accesses \texttt{HTAB}$[p]$.}
    \label{fig:partitioned-htab}
\end{figure}

\subsection{Shared-Memory Organization}

Each constraint domain is represented by one or more logical indexes backed
by shared hash structures. For example, the \texttt{Person} hierarchy domain
shown in Figure~\ref{fig:prototype-architecture} maintains an identity index
keyed by \texttt{person\_id} and a uniqueness index keyed by \texttt{email}.
Each logical index is sharded across \(P\) partitions, as illustrated in
Figure~\ref{fig:partitioned-htab}. A key \(k\) is assigned to partition
$
p = h(k) \bmod P.
$
Each partition is implemented as an independent PostgreSQL \texttt{HTAB} and
is paired with an \texttt{LWLock}. An operation first acquires
\(\texttt{LWLock}[p]\) and then invokes \texttt{hash\_search} on
\(\texttt{HTAB}[p]\). The selected \texttt{HTAB} subsequently applies its
internal hashing mechanism to locate the appropriate bucket and entry chain.
Because different partitions have independent hash tables and locks,
operations on keys assigned to different partitions can proceed concurrently.
A constraint domain containing multiple logical indexes maintains a separate
\texttt{HTAB} partition array and \texttt{LWLock} array for each index. These
partition-level \texttt{LWLock}s are held only while accessing or modifying
the physical shared-memory structures. They are distinct from the
transaction-scoped advisory locks derived from \(L_{\Gamma}\), which are held
until commit or abort and serialize transactions that conflict on conceptual
identities or constrained values.

\subsection{Transaction Integration and SQL Interface}

\vspace{4pt}
\noindent
\textbf{\ul{Logical locking.}}
Logical conflicts are coordinated using PostgreSQL transaction-scoped advisory
locks derived from conceptual identities and constrained values. Let \(S(x)\)
and \(X(x)\) denote shared and exclusive transaction-duration locks,
respectively, on logical lock identifier \(x\). Let \(k_E\) be an entity key
and \(d\) a weak-entity discriminator. The functions
\(\ell_E(k_E)\) and \(\ell_W(k_E,d)\) deterministically map entity and
weak-entity identities, respectively, to PostgreSQL advisory-lock identifiers.
An operation that creates or removes an entity acquires
$
X\bigl(\ell_E(k_E)\bigr).
$
Inserting a weak-entity instance instead acquires
$
S\bigl(\ell_E(k_E)\bigr)
\quad\text{and}\quad
X\bigl(\ell_W(k_E,d)\bigr).
$
The shared lock on the owner identity prevents the owner from being removed
while the dependent object is inserted. The exclusive lock on
\((k_E,d)\) prevents concurrent operations from creating, updating, or
removing the same weak-entity identity. An operation requiring multiple locks
deduplicates its lock identifiers and acquires the remaining locks according
to a global total order, preventing deadlocks. PostgreSQL retains these locks
until the transaction commits or aborts.

\vspace{4pt}
\noindent
\textbf{\ul{SQL interface.}}
Operations for each constraint domain are exposed through a uniform set of
SQL-callable functions. Identity domains provide \texttt{reserve},
\texttt{require}, \texttt{update}, and \texttt{remove} operations.
\texttt{Reserve} claims a new conceptual identity and rejects the operation if
that identity or another constrained value is already present.
\texttt{Require} verifies that a referenced conceptual object exists and
acquires a shared logical lock that prevents its concurrent removal or
incompatible modification. \texttt{Update} atomically replaces constrained
values while maintaining all affected index entries, and \texttt{remove}
removes the entries associated with a conceptual object.

Cardinality domains provide operations for creating and removing endpoint-count
records and for reserving, releasing, or moving relationship occurrences.
These compound operations validate the required endpoints and cardinality
bounds before atomically updating the affected counters. Administrative
operations include \texttt{load} and \texttt{clear} for reconstruction,
\texttt{get} and \texttt{count} for inspection, and
\texttt{reserved\_bytes} for memory accounting.

\vspace{4pt}
\noindent
\textbf{\ul{Reconstruction and normal enforcement.}}
For example, the \texttt{Person} identity domain is reconstructed from a
consistent scan of all physical relations in the hierarchy while normal writes
are quiesced:
\begin{verbatim}
CREATE EXTENSION IF NOT EXISTS inmem_indexes;
SELECT inmem_clear_person_identities();
SELECT inmem_load_person_identity(person_id, email)
FROM (
    SELECT person_id, email
    FROM relation_0
    UNION ALL
    SELECT student_id AS person_id, email
    FROM relation_1
    UNION ALL
    SELECT instructor_id AS person_id, email
    FROM relation_2
) AS person_extent;
\end{verbatim}
During normal execution, the reservation operation and the corresponding
relational mutation execute within the same PostgreSQL transaction:
\begin{verbatim}
BEGIN;
SELECT inmem_reserve_person_identity(
    1001, 'student1001@umd.edu'
);
INSERT INTO relation_1(student_id, email, major)
VALUES (1001, 'student1001@umd.edu',
        'Computer Science');
COMMIT;
\end{verbatim}
The reservation atomically claims both the hierarchy-wide identity and, when
present, the unique email value. A constraint violation raises an error and
aborts the transaction. Transaction-scoped locks and undo processing ensure
that the relational and auxiliary states remain consistent after either commit
or abort.

\vspace{4pt}
\noindent
\textbf{\ul{Undo and transaction callbacks.}}
For each conceptual update, the SQL first invokes the required
logical-lock and constraint-index operations and then performs the relational
mutation within the same PostgreSQL transaction. Each tentative
constraint-index mutation appends an undo record containing its inverse
operation to a unified per-transaction LIFO journal. On commit, a transaction
callback discards the journal. On abort, the callback replays its records in
reverse order, restoring the previous auxiliary state while the
transaction-scoped logical locks are still held.
The discussion here concerns whole-transaction commit and abort.
Savepoint and subtransaction rollback require additional handling and
are not established by this description.

\subsection{Recovery}

Constraint indexes are derived structures and are not independently WAL
logged (this is 
a major difference from the {\em shadow table} approach). After PostgreSQL recovers the durable relational representation,
initialization procedures scan the participating relations and
reconstruct the auxiliary indexes. Normal execution therefore avoids
additional persistent tuples, buffer management, WAL logging, and persistent
index maintenance for constraint state. Reconstruction fails if it encounters
a duplicate that violates the corresponding constraint.

\begin{table*}[t]
\centering
\small
\setlength{\tabcolsep}{4pt}
\caption{Throughput (TPS) at 256 concurrent clients. 
Speedup is relative to the better of correctness-preserving \emph{Shadow} and
\emph{Shadow-Trigger}; retention is the percentage of no-constraint throughput
achieved by the constraint index.}
\label{tab:eval-summary}
\begin{tabular}{lcccccccc}
\toprule
Workload & No const. & Query & Query-trigger & Shadow & Shadow-trigger
& Constr.\ idx. & Speedup & Retention \\
\midrule

\multicolumn{9}{l}{\emph{Distributed hierarchy}}\\
Person insert
& 87.7K & 31.1K & 61.9K & 55.1K & 57.8K & 79.6K
& $1.38\times$ & 90.9\% \\

Person mixed operations
& 69.6K & 32.0K & 44.4K & 47.6K & 44.4K & 60.5K
& $1.27\times$ & 87.0\% \\

\addlinespace
\multicolumn{9}{l}{\emph{Folded relationship}}\\
\texttt{Belongs\_to} insert
& 46.2K & 23.7K & 34.2K & 34.5K & 37.6K & 40.9K
& $1.09\times$ & 88.5\% \\

\texttt{Belongs\_to} mixed operations
& 57.1K & 30.8K & 39.2K & 33.5K & 40.8K & 49.5K
& $1.21\times$ & 86.7\% \\

\texttt{Belongs\_to} insert, single table
& 96.6K & 38.4K & 64.0K & 65.9K & N/A & 82.7K
& $1.25\times$ & 85.6\% \\

\quad\parbox[t]{0.25\textwidth}{\emph{vs.\ native PostgreSQL \texttt{UNIQUE}}}
& 96.6K & 38.4K & 64.0K & 65.9K & N/A & 68.0K
& $1.03\times$ & 70.4\% \\

\addlinespace
\multicolumn{9}{l}{\emph{Folded weak entities}}\\
Folded \texttt{Address} insert
& 41.9K & 23.7K & 35.1K & 30.2K & 33.4K & 41.3K
& $1.24\times$ & 98.7\% \\

Folded \texttt{IdCard} insert
& 40.5K & 5.3K & 4.8K & 24.9K & 29.6K & 32.9K
& $1.11\times$ & 81.4\% \\

\texttt{IdCard.card\_no} update
& 30.9K & 4.3K & 5.1K & 19.8K & 21.2K & 28.5K
& $1.35\times$ & 92.1\% \\

\addlinespace
\multicolumn{9}{l}{\emph{Beyond key-oriented constraints}}\\
Student insert with FD
& 76.7K & 31.2K & 48.2K & 37.4K & 48.6K & 71.5K
& $1.47\times$ & 93.2\% \\

Advisor update, cardinality
& 49.2K & 31.6K & 37.6K & 18.1K & 25.2K & 38.7K
& $1.54\times$ & 78.6\% \\

contributes\_to insert, cardinality
& 91.9K & 17.3K & 42.8K & 32.9K & 36.6K & 88.1K
& $2.41\times$ & 95.9\% \\

\addlinespace
\multicolumn{9}{l}{\emph{End-to-end}}\\
Person, IdCard, \texttt{Belongs\_to} mixed
& 52.6K & 12.6K & 15.0K & 37.1K & 34.9K & 44.4K
& $1.20\times$ & 84.3\% \\

\bottomrule
\end{tabular}
\end{table*}

\section{Experimental Evaluation}
\label{sec:evaluation}

We evaluate the extension integrated into Postgres on constraints distributed across inheritance
hierarchies, folded relationships, and folded weak entities, along with
an end-to-end workload that composes several constraint indexes in one
transaction. 

In addition to the entities and relationships shown in
Figure~\ref{fig:er_example}, the conceptual schema used in our experiments
contains a \texttt{Department} entity, the weak entities \texttt{Address}
and \texttt{IdCard} owned by \texttt{Person}, and a many-to-one
\texttt{Belongs\_to} relationship from \texttt{Person} to
\texttt{Department}.
In addition to its conceptual primary key \texttt{person\_id},
\texttt{Person} has a unique \texttt{email} attribute and the functional
dependency
\(\texttt{role} \rightarrow \texttt{access\_profile}\).
Weak-entity identities are
$
(\mathit{person\_id},\mathit{address\_id})
$
and
$
(\mathit{person\_id},\mathit{card\_id}),
$
and \texttt{IdCard.card\_no} is globally unique. The
\texttt{Belongs\_to} relationship also has a unique \texttt{index}
attribute.
The schema includes several cardinality constraints. An
\texttt{Instructor} may advise at most five \texttt{Student} instances.
For the many-to-many \texttt{contributes\_to} relationship, each
\texttt{Person} may contribute to at most 20 \texttt{Content} instances,
while each \texttt{Content} instance may have at most 10 contributing
\texttt{Person} instances.

Under the primary relational mapping, \texttt{Person}, \texttt{Student}, and
\texttt{Instructor} are represented by separate relations.
\texttt{Department} and the many-to-many \texttt{contributes\_to}
relationship are represented by separate relations. \texttt{Address} and
\texttt{IdCard} instances are folded into their owner tuples as arrays of
composite values and are consequently distributed across the three hierarchy
relations. The many-to-one \texttt{Belongs\_to} and \texttt{Advisor}
relationships are represented by columns folded into the appropriate
hierarchy tuples.

For the \texttt{Belongs\_to} experiments, we additionally evaluate an
alternative mapping in which the relationship is stored in a separate
relation (single table), while the \texttt{Person} hierarchy remains distributed across
three relations. Uniqueness of the relationship's \texttt{index}
attribute is therefore local to one relation, but referential integrity
becomes nonlocal: the person endpoint may reference a tuple in any of
the three hierarchy relations, which a conventional
\texttt{FOREIGN KEY} cannot express.
We consider two \emph{constraint-index} enforcement configurations under this mapping. In one configuration, constraint indexes enforce both
referential integrity and uniqueness of the \texttt{index} attribute.
In the \emph{hybrid} configuration, a constraint index enforces
referential integrity, while PostgreSQL's native \texttt{UNIQUE}
constraint enforces uniqueness of \texttt{index}. 
Because both configurations use the same constraint-index
mechanism for referential integrity, their comparison isolates the overhead
of constraint-index uniqueness relative to native declarative uniqueness.

We use PostgreSQL~16 in a client-server
configuration on two CloudLab \texttt{c220g1} nodes connected by
10 Gbps Ethernet~\cite{Duplyakin+:ATC19}. Each node has
128\,GB of memory and a 480\,GB SSD. The initialized database
contains approximately 5.7M entity instances and several million
folded weak-entity and relationship values. Each client worker maintains a dedicated
PostgreSQL connection and issues requests in a closed loop. 
We compare six enforcement configurations
over insert, update, mixed, and end-to-end workloads, for
1 to 1024 clients:
\begin{itemize}
    \item \emph{No const.} executes the workload without constraint
    enforcement and provides the upper-bound throughput reference.
    \item \emph{Query} explicitly validates constraints with
    \texttt{SELECT} queries inside the transaction that performs the
    corresponding modification.
    \item \emph{Query-trigger} performs query-based validation through
    triggers on the source relations, invoking the checks when a source
    modification can affect a constraint.
    \item \emph{Shadow} maintains auxiliary relational tables that
    represent constraint-relevant state and enforces the constraints
    using those tables.
    \item \emph{Shadow-trigger} uses triggers on the maintained auxiliary
    tables to enforce the constraints.
    \item \emph{Constr.\ idx.} uses the proposed constraint indexes.
\end{itemize}
Table~\ref{tab:eval-summary}
reports the throughputs for 256 clients, which we observed to be close to the saturation points. 
The mixed workloads consist of 40\% inserts, 30\% updates, 20\% deletes, and
10\% reads. In the end-to-end workload, 40\% of requests operate on
\texttt{Person}, 30\% on \texttt{Belongs\_to}, and 30\% on
\texttt{IdCard}. Within each operation class, requests follow the same
distribution: 40\% inserts, 30\% updates, 20\% deletes, and 10\% reads.
For \emph{Constr.\ idx.}, retention is throughput as a percentage of
\emph{No const.}, and speedup is relative to the faster correctness-preserving
baseline, \emph{Shadow} or \emph{Shadow-trigger}. The \emph{Query} and
\emph{Query-trigger} baselines run at \texttt{READ COMMITTED}; their snapshot-based checks can admit violations under concurrent execution.
For \texttt{Belongs\_to} stored in a single relation, the \emph{Shadow} configuration uses native UNIQUE to enforce index uniqueness and a
conventional foreign key from the person endpoint to a maintained
hierarchy-wide identity shadow table. These declarative constraints
make a separate \emph{Shadow-trigger} configuration unnecessary, so the
corresponding table entries are N/A.

The hybrid row retains constraint-index referential integrity and uses
native \texttt{UNIQUE}; its speedup is also measured against the
\emph{Shadow} baseline.

\begin{figure*}[t]
    \centering

    \begin{subfigure}[t]{0.49\textwidth}
        \centering
        \includegraphics[
            width=\linewidth,
            trim={0 0 0 20},
            clip
        ]{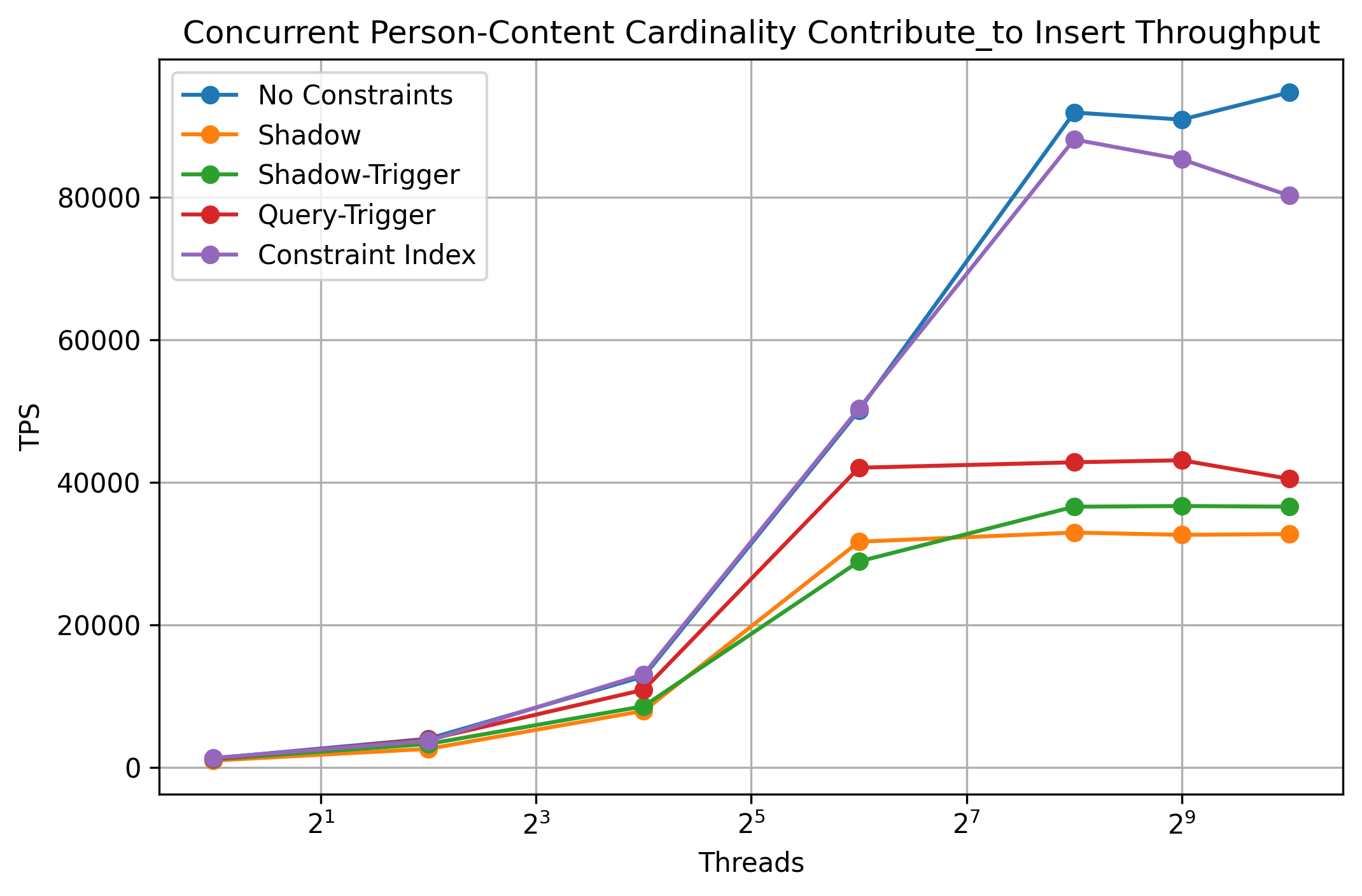}
        \caption{Throughput.}
        \label{fig:person-content-cardinality-throughput}
    \end{subfigure}
    \hfill
    \begin{subfigure}[t]{0.49\textwidth}
        \centering
        \includegraphics[
            width=\linewidth,
            trim={0 0 0 20},
            clip
        ]{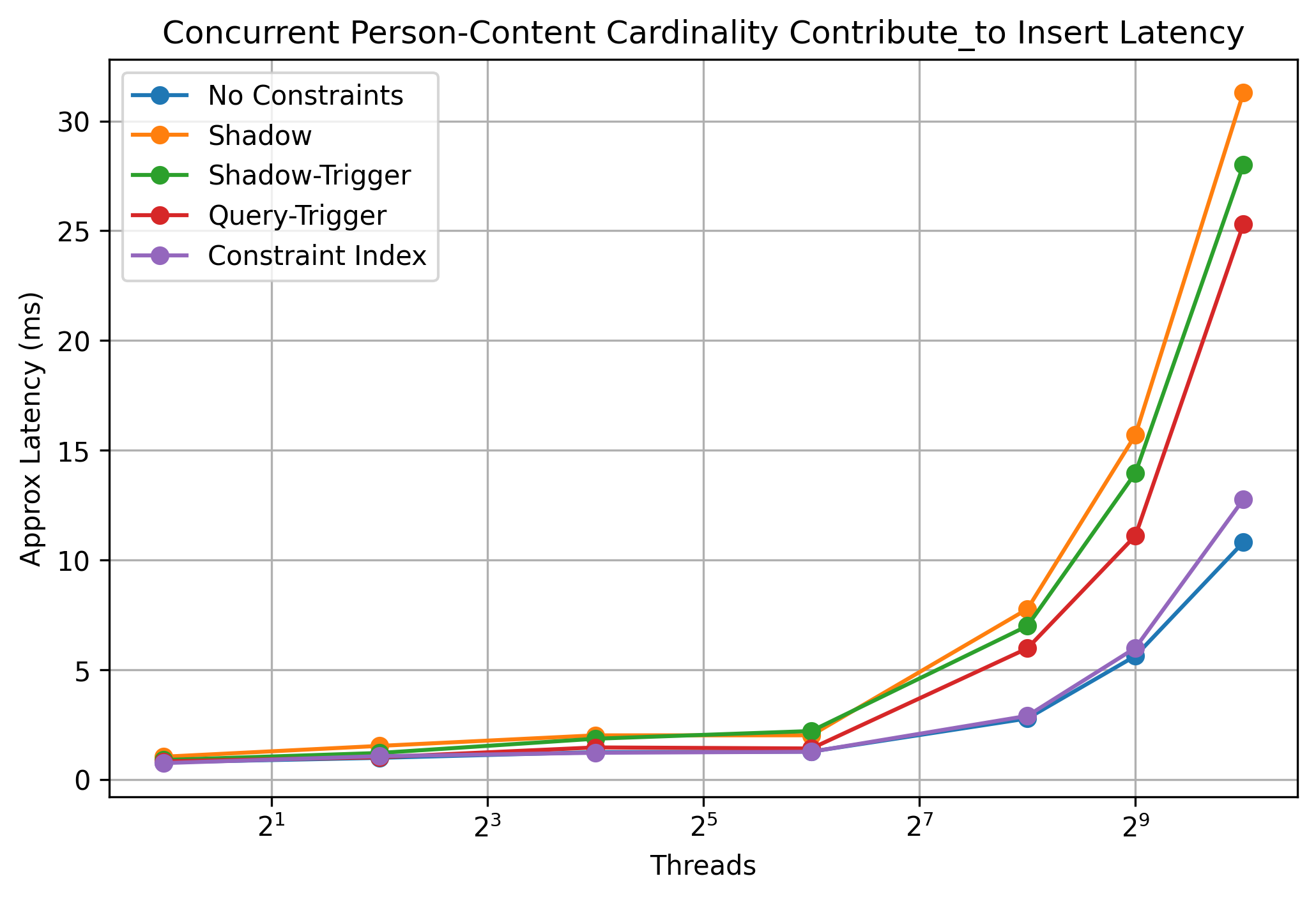}
        \caption{Approximate latency.}
        \label{fig:person-content-cardinality-latency}
    \end{subfigure}

    \vspace{-5pt}
    \caption{Scalability of \texttt{contributes\_to} inserts under the
    two-sided \texttt{Person}--\texttt{Content} cardinality constraint.}
    \vspace{-5pt}
    \label{fig:person-content-cardinality}
\end{figure*}

\vspace{4pt}\noindent
\textbf{Cost of enforcement.}
As we can see, constraint indexes retain between 81.4\% and 98.7\% of unconstrained
throughput across the key-oriented workloads using constraint indexes
for all evaluated constraints in Table~\ref{tab:eval-summary}.
Retention is highest when enforcement
requires a single lookup path, such as folded \texttt{Address}
insertion, and lowest when one logical operation maintains several
structures. In particular, \texttt{IdCard} operations maintain a
composite identity index, a unique \texttt{card\_no} index, and an
owner-to-dependent index. Even in this case, the results indicate that
cross-relation constraints are inexpensive enough to remain enabled.

\vspace{4pt}\noindent
\textbf{Comparison with relational enforcement.}
Constraint indexes outperform the best trigger or shadow-table baseline
on every key-oriented workload, by $1.09\times$ to $1.38\times$,
when constraint indexes enforce all evaluated constraints. The largest
gain is for \texttt{Person} insertion. Updating
\texttt{IdCard.card\_no} achieves a $1.35\times$ improvement:
relational enforcement separately maintains
extracted identity and unique-value state, whereas the constraint index
updates one conceptual object and its lookup entries directly. The
smallest gain is for simple \texttt{Belongs\_to} insertion, where the
underlying owner-tuple update dominates execution. 
For \texttt{Belongs\_to} inserts under the mapping that stores the
relationship in a separate relation (single table), constraint indexes enforcing both referential integrity and
uniqueness achieve 82.7K TPS. The hybrid configuration, which retains
constraint-index enforcement of referential integrity but uses
PostgreSQL's native \texttt{UNIQUE} constraint for the \texttt{index}
attribute, achieves 68.0K TPS. Thus, the constraint-index configuration
achieves $1.22\times$ the throughput of the hybrid configuration.
Because both configurations use the same referential-integrity mechanism,
this comparison isolates the performance effect of the uniqueness
implementation.

Additional experiments (not shown) under
the single-table hierarchy mapping indicate
that the benefit does not arise
only from avoiding cross-relation checks; constraint indexes also avoid
persistently materializing keys extracted from nested values.
We note that since query and query-trigger baselines run at READ COMMITTED, they are not sound in several of these cases; 
using SERIALIZABLE with retries or explicit locking would add synchronization costs, which we have not measured here.

Explicit validation queries perform substantially worse. For example,
\texttt{IdCard} insertion reaches 5.3K TPS with query-based checking,
compared with 32.9K TPS using constraint indexes. This is also the
approach most similar to the SQL generated by unaided LLM writers, and, 
without additional synchronization, both
query-based and trigger-based checking remain vulnerable to the
write-skew anomaly described in Section~\ref{sec:triggers}.

\vspace{4pt}\noindent
\textbf{Composition and generality.}
The end-to-end workload combines hierarchy-wide identity and email
uniqueness, folded-relationship constraints, folded weak-entity identity
and uniqueness, and cascading deletions that update several constraint
indexes in one transaction. Constraint indexes achieve 44.4K TPS,
compared with 37.1K for relational enforcement and 52.6K without
constraints: a $1.20\times$ improvement while retaining 84.3\% of
unconstrained throughput. 

We also experimented with more complex constraints.
For \texttt{Student} inserts subject to functional dependency
\(\texttt{role} \rightarrow \texttt{access\_profile}\), constraint indexes
improve throughput by \(1.47\times\) over the strongest alternative
enforcement mechanism while retaining 93.2\% of unconstrained throughput.
For inserts into \texttt{contributes\_to} under the two-sided cardinality
constraint, constraint indexes provide a \(2.41\times\) improvement while
retaining 95.9\% of unconstrained throughput. Figure~\ref{fig:person-content-cardinality}
shows the throughput and latency comparison as concurrency increases.
In this case, the shadow-table baseline maintained two auxiliary count tables—one for each 
endpoint hierarchy—with a check constraint on each count column to enforce the corresponding cardinality bound.
Inserting a relationship therefore requires validating its endpoints and
transactionally updating both count tables. 
We view these results as evidence that the compilation recipe
extends beyond keys, rather than as a general solution for arbitrary
constraints.

Finally, the prototype remains a PostgreSQL extension rather than a
native engine feature, primarily to make it easier to experiment with 
different alternatives. Each operation crosses a SQL-callable function
boundary, uses general-purpose advisory locks, and is invisible to the
planner and executor. Native integration could eliminate redundant
checks, batch or fuse enforcement with relational updates, and use
purpose-built lock modes. 

\section{Conclusion \& Research Challenges}
\label{sec:conclusion}
\label{sec:challenges}

Leaving integrity constraints to the application layer made sense when application code was durable, reviewed, and written by people who understood the invariants, an assumption that no longer holds for all writers.
Constraints are increasingly the specification of valid data that outlives the code operating on it, and the database is the last component able to enforce them deterministically. 
Constraint indexes demonstrate that one substantial class of constraints generated by conceptual models, long assumed too expensive to enforce, can instead be compiled into transactional 
access paths at a cost low enough to leave switched on. Treating integrity enforcement as a physical-design problem, with enforcement structures compiled and selected alongside 
the relational schema is, we believe, the way to close the remaining gap.

The first challenge is to automate and generalize the construction of
constraint indexes and integrate them more deeply into the database engine. Prior work has largely incrementalized constraint \emph{checking}, while treating synchronization separately.
Our premise is that the two should be handled together, because data structures that make a check cheap are also expressive enough to manage the conflicts because of it.
Our experiments extend this approach beyond keys to cardinality bounds
and functional dependencies. Applying it to conditional uniqueness,
participation constraints, interval disjointness, and other aggregate
conditions remains an open direction.
Its reach depends largely on whether the auxiliary state can be maintained incrementally under insertion and deletion. 
Arbitrary assertions requiring unrestricted query evaluation may remain outside the approach, although the recent work
on incremental view maintenance~\cite{budiu2023dbsp} may help unify these. 
A complete system must also choose among alternatives that trade storage, maintenance, validation cost, and contention, and identify opportunities for sharing across constraints. 
An interesting research direction here is also to explore approximate data structures such as Bloom filters or sketches, to further improve performance.

The second challenge is governance: deciding how each constraint should be enforced and how that decision changes over
time. This includes not only the enforcement structure, but also whether checking occurs per statement, at commit,
or asynchronously, and whether a violation aborts, quarantines, or merely raises an alert. Some constraints
inherently create broad contention or require expensive state, in which case weaker enforcement may be preferable to
an elaborate and expensive index. Such flexibility is also essential for the constraint lifecycle. Candidate
constraints should be deployable provisionally against live traffic, promoted to blocking enforcement when evidence
supports them, and demoted or dropped when it does not. This becomes especially important when constraints are
inferred from application code, workloads, or existing data rather than supplied by a conceptual schema: inferred
constraints are necessarily uncertain, and provisional enforcement can make it safe to propose them. Open questions
include how promotion evidence should be evaluated, how pre-existing violations should be treated, and how
provisional enforcement can remain proportional to a candidate constraint's uncertain value.

Finally, the enforcement boundary must support the writers (like AI agents) that motivate this work. Writers should be able to
test a proposed transaction without committing it, and a rejection should return both the violated constraint and a
witness identifying the conflicting values, so that a repair loop can converge on the intended correction. Constraints
should also be exposed in machine-readable form. Our probe shows, however, that description alone is insufficient:
models given constraints as text produced widely divergent and frequently unsafe code. The database must
therefore provide both semantic context to the writer and enforcement that remains independent of the writer's behavior.

\section{AI-Generated Content Acknowledgement}
The authors used ChatGPT (OpenAI) and Claude (Anthropic) to assist
with refining and editing portions of the manuscript for clarity
and presentation. These tools also assisted with debugging the
PostgreSQL extension, setting up the concurrency experiments, and generating the dataset for the LLM experiment.

\bibliographystyle{ACM-Reference-Format}
\bibliography{sample-base}

\end{document}